\documentclass[twocolumn,showpacs,preprintnumbers,nofootinbib,draft,tightenlines,prd,aps,
floats,amsmath,amssymb,superscriptaddress]{revtex4} 

\usepackage{amsmath,amssymb,latexsym}
\usepackage{bm}

\newcommand{\al}{\alpha}
\newcommand{\be}{\beta}
\newcommand{\om}{\omega}

\newcommand{\si}{\sigma}
\newcommand{\ga}{\gamma}
\newcommand{\de}{\delta}
\newcommand{\nn}{\nonumber}
\newcommand{\cP}{{\cal{P}}}
\newcommand{\cB}{{\cal{B}}}

\begin{document}


\preprint{hep-th/0208188}

\title{On T-Duality in  Brane Gas Cosmology}

\author{Timon Boehm }


\email[Electronic address: ]{Timon.Boehm@physics.unige.ch}

\affiliation{D\'epartement de Physique Th\'eorique, Universit\'e
de Gen\`eve, 24 quai E. Ansermet, CH-1211 Geneva 4, Switzerland.}

\author{Robert Brandenberger}

\email[Electronic address: ]{rhb@het.brown.edu}

\affiliation{Institut d'Astrophysique de Paris, 98bis Blvd. Arago,
F-75014 Paris, France,
\\and\\
Physics Department, Brown University, Providence, RI 02912, USA.}

\date{\today}



\begin{abstract}
In the context of homogeneous and isotropic
superstring cosmology, the T-duality symmetry
of string theory has been used to argue that
for a background space-time described by dilaton gravity with
strings as matter sources, the cosmological evolution of the
Universe will be nonsingular. In this Letter we discuss how
T-duality extends to brane gas cosmology, an approximation in
which the background space-time is again described by dilaton
gravity with a gas of branes as a matter source. We conclude that
the arguments for non-singular cosmological evolution remain valid.
\end{abstract}


\maketitle 


\section{Introduction}

In \cite{Brandenberger:1989aj} it was suggested that due to a new
string theory-specific symmetry called T-duality, string theory
has the potential to resolve the initial singularity problem of
Standard Big Bang cosmology, a singularity which also plagues
scalar field-driven inflationary cosmology
\cite{Borde:1994xh,Borde:2001nh}.

The framework of \cite{Brandenberger:1989aj} was based on an
approximation in which the mathematical background space-time is
described by the equations of dilaton gravity (see
\cite{Tseytlin:1992xk,Veneziano:1991ek}), with the matter source
consisting of a gas of strings. The background spatial sections
were assumed to be toroidal such as to admit one-cycles. Thus, the
degrees of freedom of the string gas consist of winding modes in
addition to the momentum modes and the oscillatory modes. Then,
both momentum and winding numbers take on discrete values, and the
energy spectrum of the theory is invariant under inversion of the
radii of the torus, i.e. $R \rightarrow \al'/R$, where $\al^{'
1/2}$ is the string length $l_s$. The mass of a state with
momentum and winding numbers $n$ and $\omega$, respectively, in a
compact space of radius $R$, is the same as that of a state with
momentum and winding numbers $\omega$ and $n$, respectively, in
the space of radius $\al'/ R$. This symmetry was used to argue
\cite{Brandenberger:1989aj} that as the radii of the torus
decrease to very low values, no physical singularities will occur.
Firstly, under the assumption of thermal equilibrium, the
temperature of a string gas at radius $R$ will be equal to the
temperature of the string gas at radius $\al'/ R$. Secondly, any
process computed for strings on a space with radius $R$, is
identical to a dual process computed for strings on a  space with
radius $\al'/ R$. Therefore there exists a `minimal' radius
$\al^{' 1/2}$ in the sense that physics on length scales below
this radius can equally well be described by physics on length
scales larger than that.

Since the work of \cite{Brandenberger:1989aj} our knowledge of
string theory has evolved in important ways. In particular, it has
been realized \cite{Polchinski:1995mt} that string theory must
contain degrees of freedom other than the perturbative string
degrees of freedom used in \cite{Brandenberger:1989aj}. These new
degrees of freedom are Dp-branes of various dimensionalities
(depending on which string theory one is considering). Since the
T-duality symmetry was used in an essential way (see e.g.
\cite{Polchinski:1998rr}) to arrive at the existence of Dp-branes
(p-branes for short in the following), it is clear that T-duality
symmetry will extend to a cosmological scenario including
p-branes. However, since a T-duality transformation changes the
dimensionality of branes, it is useful to explicitly verify that
the arguments of \cite{Brandenberger:1989aj} for a non-singular
cosmological evolution carry over when the gas of perturbative
string modes is generalized to a gas of branes. A model for
superstring cosmology in which the background space-time is
described (as in \cite{Tseytlin:1992xk,Veneziano:1991ek}) by
dilaton gravity, and the matter source is a gas of branes, has
recently been studied under the name of ``brane gas cosmology''
\cite{Alexander:2000xv,Brandenberger:2001kj} (see also
\cite{Easson:2001fy,Easther:2002mi} for extensions to backgrounds
which are not toroidal, and \cite{Watson:2002nx} for an extension
to an anisotropic background).

In this Letter, we establish the explicit action of T-duality in
the context of brane gas cosmology on a toroidal background. For a
solution of the background geometry appropriate for cosmological
considerations in which the radii of the torus are decreasing from
large to small values as we go back in time, we must consider
T-dualizing in all spatial dimensions. We demonstrate that the
mass spectrum of branes remains invariant under this action. Thus,
if the background dynamics are adiabatic, then the temperature of
the brane gas will be invariant under the change $R \rightarrow
{\al'/ R}$, i.e.
\begin{equation}
T(R) \, = \, T \left(\frac{\al'}{R}\right) \, ,
\end{equation}
thus demonstrating that superstring cosmology can avoid the
temperature singularity problem of standard and inflationary
cosmology. In the appendix we make further remarks on why we expect
brane gas cosmology to be non-singular. Another crucial assumption is that
the string coupling constant $g$ is small (compared to one) such that
back-reactions of the string and brane gas on the curvature of space-time can be
ignored.  Similarly, our results can be used to show that the
arguments for the existence of a minimal physical length given in
\cite{Brandenberger:1989aj} extend to brane gas cos\-mo\-logy.

The outline of this Letter is as follows. In the following section
we give a brief review of brane gas cosmology. In Section
\ref{sec:energy} we (partially re-)derive the energy, the
momentum, and the pressure for  p-branes.
Next, we review the action of T-duality on winding states of
p-branes. The main section of this Letter is Section
\ref{sec:spectrum} in which we show that the mass spectrum of a
p-brane gas of superstring theory is invariant under T-duality.
Section \ref{sec:cosmology} contains a discussion of some
implications of the result, and conclusions. 


\section{Review of Brane Gas Cosmology}
\label{sec:review}

As already mentioned in the Introduction, the framework of brane
gas cosmology consists of a homogeneous and isotropic background
of dilaton gravity coupled to a gas of p-branes as a matter
source. We are living in the bulk \footnote{In this sense, brane
gas cosmology is completely different in ideology than brane world
scenarios in which it is assumed (in general without any dynamical
explanation) that we live on a specific brane embedded in a warped
bulk space-time. From the point of view of heterotic M-theory
\cite{Horava:1996qa}, our considerations should be viewed as
applying to the 10 dimensional orbifold space-time on which we
live.}.

The initial conditions in the early Universe are `conservative'
and `democratic'; conservative in the sense that they are close to
the initial conditions assumed to hold in standard big bang
cosmology (i.e. a hot dense gas of matter), democratic in the
sense that all 9 spatial dimensions of critical string theory are
considered on an equal basis\footnote{For consistency, critical
superstring theories need a 10 dimensional target space-time which
is in apparent contradiction with the observed four. Usually it is
assumed that six dimensions are compactified from the outset due
to some unknown physics. However, following the usual approach in
cosmology it seems more natural that initially all nine spatial
dimensions were compact and small, and that three of them have
grown large by a dynamical {\em decompactification}
process.  A corresponding scenario was originally proposed in 
\cite{Brandenberger:1989aj}.}.
Thus, matter is
taken to be a gas of p-branes of all allowed values of $p$ in
thermal equilibrium. In par\-ti\-cu\-lar, all modes of the branes
are excited, including the winding modes.

The background space-time is taken to be ${\mathbb{R}} \times T^9$
where $T^9$ denotes a nine-torus. The key feature of $T^9$ which
is used in the analysis is the fact that it admits one-cycles
which makes it possible for closed strings to have conserved
winding numbers \footnote{Recently, the scenario of
\cite{Brandenberger:1989aj,Alexander:2000xv} was generalized
\cite{Easson:2001fy,Easther:2002mi} to spatial backgrounds such as
Calabi-Yau manifolds which admit 2-cycles but no 1-cycles.}. It is
also assumed that the initial radius in each toroidal direction is
the same, and comparable to the self-dual radius $\al^{'1/2}$.
Initially, all directions are expanding isotropically with $R >
\al^{'1/2}$ (the extension to anisotropic initial conditions has
recently been considered in \cite{Watson:2002nx}).

As shown in this Letter, as a consequence of T-duality symmetry,
brane gas cosmology provides a background evolution without
cosmological singularities. The scenario also provides a possible
dynamical explanation for why only three spatial dimensions can
become large. Winding modes (and thus T-duality) play a crucial
role in the argument. Let us first focus on the winding modes of
fundamental strings \cite{Brandenberger:1989aj}.

The winding and anti-winding modes $\om$ and $\bar{\om}$
of the strings are initially in thermal
equilibrium with the other states in the string gas. Thermal
equilibrium is maintained by the process
\begin{equation} \label{eq:equilibrium}
    \om + \bar{\om} \rightleftarrows \rm{loops, radiation}.
\end{equation}
When strings cross each other, they can intercommute such that a
winding and an anti-winding mode annihilate, producing fundamental
string loops or radiation without winding number. This process is
analogous to infinite cosmic strings intersecting and producing
cosmic string loops and radiation during their interaction (see
e.g. \cite{VilenkinShellard,Brandenberger:1994by} for reviews of
cosmic string dynamics).

As the spatial sections continue to expand, matter degrees of
freedom will gradually fall out of equilibrium. In the context of
string gas cosmology with $R > \al^{'1/2}$, the winding strings
are the heaviest objects and will hence fall out of equilibrium
first. Since the energy of a winding mode is proportional to $R$,
Newtonian intuition would imply that the presence of winding modes
would prevent further expansion. This is contrary to what would be
obtained by using the Einstein equations. However, the equations
of dilaton gravity yield a similar result to what is obtained from
Newtonian intuition \cite{Tseytlin:1992xk}: the presence of
winding modes (with negative pressure) acts as a confining
potential for the scale factor.

As long as winding modes are in thermal equilibrium, the total
energy can be minimized  by transferring it to momentum or
oscillatory  modes (of the fundamental string). Thereby, the
number of winding modes decreases, and the expansion can go on.
However, if the winding modes fall out of equilibrium, such that
there is a large number of them left, the expansion is slowed down
and eventually stopped. If we now try to make $d$ of the original
9 spatial dimensions much larger than the string scale, then an
obstruction is encountered if $d > 3$: in this case the
probability for crossing and therefore for equilibrating according
to the process (\ref{eq:equilibrium}) is zero. On the other hand,
in a three-dimensional subspace of the nine-torus, two strings
will generically meet. Therefore, the winding modes can
annihilate, thermal equilibrium can be maintained, and, since the
decay modes of the winding strings have positive pressure, the
expansion can go on \footnote{Since quantum mechanically, the
thickness of the strings is given by the string length
\cite{Karliner:1988hd}, it is important for the brane gas scenario
that the initial size of the spatial sections was string scale.
Otherwise, it would always be the total dimensionality of space
which would be relevant in the classical counting argument of
\cite{Brandenberger:1989aj}, and there could be no expansion in
any direction.}.  As a result, there is no topological obstruction to
three dimensions of the torus growing large while the other six 
are staying small (of size $R \sim
\al^{'1/2}$) \;\footnote{ When making these 
considerations we have neglected the possibility that closed strings may
break up (in analogy to Hadron fragmentation in QCD). The amplitude of
this process should be investigated quantitatively.}.  
Large compact dimensions are not in contradiction with
observations if their radius is bigger than
the Hubble radius today.

It is not hard to include p-branes into the above scenario
\cite{Alexander:2000xv}. Now the initial state is a hot, dense gas
of all branes allowed in a particular theory. In particular, brane
winding modes are excited, in addition to modes corresponding to
fluctuations of the brane. Since the winding modes play the most
important role in the dynamical decompactification mechanism of
\cite{Brandenberger:1989aj,Alexander:2000xv}, we will focus our
attention on these modes. The analogous classical counting
argument as given above for strings yields the result that p-brane
winding modes can interact in at most $2p+1$ spatial dimensions.
Since for weak string coupling and for spatial sizes larger than
the self-dual radius the mass of a p-brane (with a fixed winding
number in all of its p spatial dimensions) increases as p
increases, p-branes will fall out of equilibrium earlier the
larger $p$ is. Thus, e.g. in a scenario with 2-branes, these will
fall out of equilibrium before the fundamental strings and allow
five spatial dimensions to start to grow \cite{Alexander:2000xv}.
Within these five spatial dimensions, the fundamental string
winding modes will then allow only a three-dimensional subspace to
become macroscopic.  Thus there is no topological obstruction to the
dynamical generation of a hierarchy of internal
dimensions.


\section{Energy, Momentum, and Pressure of  p-Branes}
\label{sec:energy}

This section is devoted to the derivation of physical quantities
describing the brane gas which determine the cosmological
evolution of the background space-time.

Starting from the Dirac-Born-Infeld action, we obtain expressions
for the energy and the momentum of a p-brane in $D=d+1$
dimensional space-time. We show that there is no momentum flowing
along the p tangential directions. From the energy--momentum
tensor one can also define a pressure, and hence an equation of
state, for the whole brane gas.
Even though some of the results in this section are already known,
we find it useful to give a self-consistent overview.

Let $\si=(\si^0,\si^i), i=1,\cdots,p$, denote some intrinsic
coordinates on the worldsheet of a p-brane. Its position (or
embedding) in D-dimensional space time is described by
$x^{\mu}=X^{\mu}(\si)$, where $\mu=0,\cdots,d$, and the induced
metric is $\ga_{ab}=\eta_{\mu\nu} X^\mu_{,a} X^\nu_{,b}$, where
$a,b = 0,\cdots,p$.

The Dirac-Born-Infeld action is (in the string frame)
\begin{equation} \label{eq:dbiaction}
    S_p = -T_p \int d^{p+1}\si e^{-\phi}\sqrt{-\ga} \, ,
\end{equation}
where $T_p$ denotes the tension (charge) of a p-brane,
$\ga=\mbox{det}(\ga_{ab})$, and $\phi$ is the dilaton of the
compactified theory. For our adiabatic considerations, we assume
that it is  constant (taking its asymptotic value), and absorb it
into a physical tension
\begin{equation} \label{eq:tension}
     \tau_p = e^{-\phi}T_p = \frac{T_p}{g} =
            \frac{1}{(2\pi)^p}\frac{1}{g\alpha'^{(p+1)/2}} \, ,
\end{equation}
where in the final step we have used the expression for $T_p$ (see
e.g. \cite{Polchinski:1998rr}). Note that for any p-brane $\tau_p$
goes like $1/g$, and that hence, in the weak string coupling
regime which we are considering, the branes are heavy.  The action
(\ref{eq:dbiaction}) can be written as an integral over
D-dimensional space-time
\begin{equation} \label{eq:10Daction}
    S_p=\int d^{D}x \left(-\tau_p \int d^{p+1}\si \delta^{(D)}(x^\mu-X^\mu(\si))\sqrt{-\ga}
    \right)\,.
\end{equation}
As the integration domain is a torus, both integrals are finite.

Varying the action (\ref{eq:10Daction}) with respect to the
background metric and comparing with the usual definition of the
space-time energy-momentum tensor, one obtains\footnote{In curved
backgrounds the energy-momentum tensor gets multiplied by
$\sqrt{-g}$.}
\begin{eqnarray}\label{eq:emtensor}
    &&T^{\al\be}(x^{\mu}) = \\
    &&-\tau_p \int d^{p+1}\si \delta^{(D)}(x^\mu-X^\mu(\si))\sqrt{-\ga}
    \ga^{ab} X^\al_{,a} X^\be_{,b}\,.\nn
\end{eqnarray}

The DBI action (\ref{eq:dbiaction}) is invariant under $p+1$
reparametrizations $\si \rightarrow \tilde{\si}(\si)$, and we can
use this freedom to choose
\begin{equation}\label{eq:gauge}
    \ga_{00} = -\sqrt{-\ga}\,, \qquad \ga_{0i} = 0\,.
\end{equation}
Notice that $\mbox{det}(\ga_{ij}) = -\sqrt{-\ga}$ and
$\ga^{ik}\ga_{kj}=\de^i_j$. By choosing this gauge, we do not
specify a particular embedding which will be convenient later when
treating a brane gas, where the branes have arbitrary
orientations. Furthermore, it is consistent to set $X^0 = \si^0$.

To calculate the energy $E_p$ of a p-brane, one observes that in
the gauge (\ref{eq:gauge})
\begin{eqnarray}
    &&\ga^{ab} X^0_{,a} X^0_{,b}=\ga^{00}=-\frac{1}{\sqrt{-\ga}}\nn\\
    \Rightarrow && T^{00} (x^\mu) = \tau_p \int d^{p+1}\si
    \de^{(D)}(x^\mu-X^\mu (\si))\,.
\end{eqnarray}

Writing $x^\mu = (t,x^n), n=1,\cdots,d$, and splitting the
delta-function into a product, the integral over $\si^0$ can be
carried out. The energy density of a p-brane in d spatial
dimensions is
\begin{equation}\label{eq:finalrho}
    \rho_p \equiv T^{00} (t,x^n) =  \tau_p \int d^p\si
    \de^{(d)}(x^n-X^n(t,\si^i)),
\end{equation}
and its total energy is
\begin{equation}\label{eq:Ep}
    E_p = \int d^d x \rho_p = \tau_p \int d^p \si = \tau_p
    Vol_p\,.
\end{equation}
The volume of a p-brane in its rest frame, $Vol_p$, is finite as
the brane is wrapped around a torus. Eq. (\ref{eq:Ep}) provides a
formula for the lowest mass state, $M_p = E_p$, which will be used
in section \ref{sec:spectrum}. As expected intuitively, the
minimal mass is equal to the tension times the volume of a brane.

To calculate the space-time momentum $P^n_p$ of a p-brane, one
first evaluates
\begin{eqnarray}
     &&\ga^{ab} X^0_{,a}
     X^n_{,b}=-\frac{X^n_{,0}}{\sqrt{-\ga}}\nn\\
     \Rightarrow && T^{0n} (x^\mu) = \tau_p \int d^{p+1}\si
    \de^{(D)}(x^\mu-X^\mu (\si))X^n_{,0}
 \label{eq:0n}
\end{eqnarray}
Proceeding similarly as before, the total momentum of a p-brane is
found to be
\begin{equation}
    P^n_p = \tau_p \int d^p \si \dot{X}^n(t,\si^i)\,,
\end{equation}
where the dot denotes the derivative w.r.t. time $t$. The gauge
conditions (\ref{eq:gauge}) can be written as $0 =
\ga_{0i}=\dot{X}^m X_{m,i}$, where the sum over $m=1,\cdots,d$ is
the ordinary Euclidean scalar product. This is equivalent to
saying that the (spatial) velocity vector $\dot{X}$ is
perpendicular onto each of the tangential vectors $X^m_{,i}$.
Therefore, only the transverse momentum is observable
\footnote{Note the analogy with topological defects in field
theory, where also only the transverse momentum
of the defects - here taken to be straight - is observable.}. Assuming
that the brane is a pointlike classical object w.r.t. the
transverse directions, this momentum is not quantized despite of
the compactness of space. In particular, the question whether
there might exist a T-duality correspondence between transverse
momentum modes and winding modes does not arise. Moreover, we
neglect the possibility of open strings travelling on the brane
which would in fact lead to a non-zero tangential
momentum.\footnote{A relation between brane winding modes and open
string momentum modes, quantized as $n_m/R_m$, was shown in
\cite{Sen:1996cf}.} Hence, in what follows, we focus on the zero
modes of p-branes.

Finally, the pressure $\cP_p$ of a p-brane is given by averaging
over the trace $T^m_m$. First, notice that
\begin{eqnarray}
     &&\ga^{ab} X^m_{,a}X_{m,b}\nn\\
    =&&\ga^{00}X^m_{,0} X_{m,0} + \ga^{11}\ga_{11}+ \cdots + \ga^{pp}\ga_{pp}\nn\\
          & & +2\ga^{12}\ga_{21} + \cdots + 2\ga^{1p}\ga_{p1}\nn\\
          & & +2\ga^{23}\ga_{32} + \cdots + 2\ga^{2p}\ga_{p2} + \cdots \nn\\
          & & +2\ga^{p-1,p}\ga_{p,p-1}\nn\\
    =&& -\frac{1}{\sqrt{-\ga}}X^m_{,0} X_{m,0} +
    p\,.
\end{eqnarray}
In the first step we have used that the products of the embedding
functions can be expressed in terms of the induced metric, e.g.
$\ga_{11} = X^m_{,1}X_{m,1}$, and in the second step the fact that
$\ga^{ik}\ga_{kj}=\de^i_j$. Inserting this into Eq.
(\ref{eq:emtensor}), eliminating the remaining $\sqrt{-\ga}$ by
$-\sqrt{-\ga} = \ga_{00} = -1 + X^m_{,0} X_{m,0}$, and integrating
out the $\si^0$ dependence, one finds
\begin{eqnarray}\label{eq:trace}
    &&T^m_m(t,x^n) = \\
    &&\tau_p \int d^p\si \de^{(d)}(x^n-X^n(t,\si^i))
                        [(p+1)\dot{X^m}\dot{X_m} - p]\,.\nn 
\end{eqnarray}
The quantity $\dot{X^m}\dot{X_m}$ is the squared velocity of a
point on a brane parameterized by $(t,\si^i)$. We define the mean
squared velocity of the branes in the gas by averaging over all
$\si^i$, i.e. $v^2(t)~\equiv~ \langle~\dot{X^m}\dot{X_m}\rangle$.
In the averaged trace, $\langle T^m_m \rangle$, the velocity term
can be taken out of the integral. Comparing with Eq.
(\ref{eq:finalrho}), one obtains the equation of state of a
p-brane gas
\begin{equation}
    \cP_p \equiv \frac{1}{d}\langle T^m_m \rangle = \left[\frac{p+1}{d}v^2
    -\frac{p}{d}\right]\rho_p\,.
\end{equation}
In the relativistic limit ($v^2 \rightarrow 1$) the branes behave
like ordinary relativistic particles: $\cP_p = \frac{1}{d}\rho_p$,
whereas in the non-relativistic limit ($v^2 \rightarrow 0$) $\cP_p
= -\frac{p}{d}\rho_p$. For domain walls this result was obtained
in \cite{Zeldovich:1974uw}.

The pressure $\cP_p$ and the energy density $\rho_p$ are the
source terms in the Einstein equations for the brane gas
\cite{Alexander:2000xv,Tseytlin:1992xk}.


\section{Winding states and T-duality}
\label{sec:tduality}

We briefly review some of the properties
of T-duality that are needed subsequently.

Consider a nine-torus $T^9$ with radii $(R_1,\cdots, R_9)$. Under
a T-duality transformation in n-direction
\begin{equation}\label{eq:tdualitydef}
    R_n \rightarrow R_n'=\frac{\al'}{R_n} \, ,
\end{equation}
and  all other radii stay invariant. T-duality also acts on the
dilaton (which is  constant in our case), and hence on the string
coupling constant, as
\begin{equation} \label{eq:tdualityofg}
    g \rightarrow g'=\frac{\al^{'1/2}}{R_n}g \, .
\end{equation}
Note, however, that the fundamental string length $l_s \equiv
\al^{'1/2}$ is an invariant. The transformation law
(\ref{eq:tdualityofg}) follows from the requirement that the
gravitational constant in the effective theory remains invariant
under T-duality. In general, T-duality changes also the background
geometry. However, a Minkowski background (as we are using here)
is invariant.

For a p-brane on $T^9$, a particular winding state is described by
a vector $\om = (\om_1,\cdots,\om_9)$. There are $9! / [p!
(9-p)!]$ such vectors corresponding to all possible winding
configurations. For illustration take a 2-brane on a three-torus:
it can wrap around the (12), (13), (23) directions, and hence
there are $3! / 2! = 3$ vectors $\om = (\om_1,\om_2,0),
\om=(\om_1,0,\om_3), \om=(0,\om_2,\om_3)$.

Whereas T-duality preserves the nature of a fundamental string, it
turns a p-brane into a different object. To see this consider a
brane with $p$ single windings $\om=(1,\cdots,1,0,\cdots,0)$ which
represents a p-dimensional hypersurface on which open strings end.
Along the brane the open string ends are subject to Neumann
boundary conditions. These become Dirichlet boundary conditions on
the T-dual coordinate $R'_n$ (if n denotes a tangential
direction), i.e. for each string endpoint $R'_n$ is fixed. Thus a
T-duality in a tangential direction turns a p-brane into a
(p-1)-brane. Similarly, a T-duality in an orthogonal direction
turns it into a (p+1)-brane (see e.g. \cite{Polchinski:1998rr} for
more details).

Next, consider a T-duality transformation in a direction in which
the p-brane has multiple windings $\om_n
> 1$. One obtains a number $\om_n$ of (p-1)-branes which are
equally spaced along this direction. As an example take a 1-brane
with winding $\om_1 = 2$ on a circle with radius $R_1$. This
configuration is equivalent to a 1-brane with {\em single} winding
on a circle with radius $2R_1$. T-dualizing in 1-direction gives a
single 0-brane on a circle of radius $\al'/2R_1$ which is
equivalent to two 0-branes  on a circle of radius $\al'/R_1$ (see
e.g. \cite{Hashimoto:1998pd}). Since applying a T-duality
transformation twice in the same direction yields the original
state (up to a sign in the RR field), also the inverse is true: a
number $\om_n$ of (p-1)-branes correspond to a single p-brane with
winding $\om_n$.

So far we have discussed T-duality transformations in a single
direction. For applications to isotropic brane gas cosmology we
need to consider T-dualizing in all nine spatial directions. Given
a gas of branes, $\cB$, on a nine-torus with radii $(R_1, \cdots,
R_9)$ consisting of a large number of branes of all types admitted
by a particular string theory, we want to find the corresponding
gas $\cB^*$  on the dual torus $T^*$ with radii
$(R'_1,\cdots,R'_9)$. To that end one performs a T-duality
transformation in each of the nine spatial directions. From what
we have discussed so far, it is now easy to see that a p-brane in
a winding state $\om=(\om_1,\cdots,\om_p,0,\cdots,0)$ is mapped
into a number $\om_1 \cdots \om_p$ of (9-p)-branes, each of which
is in a state $\om^*=(0,\cdots,0,1,\dots,1)$. The (9-p)-brane
wraps in the (9-p) directions orthogonal to the original p-brane.
It is clear that the above considerations hold for any winding
configuration.

After these preparatory steps, we now turn to the main part of this
Letter.


\section{Mass spectra and T-duality}
\label{sec:spectrum}

\subsection{Masses of p-branes with single winding}

In this section we show that each mass state in a brane gas $\cB$
has a corresponding state with equal mass in the brane gas
$\cB^*$. Based on type IIA superstring theory we take $\cB$ to
consist of 0, 2, 4, 6 and 8 branes. Then, by the discussion in the
preceding section, the brane gas $\cB^*$ contains 9, 7, 5, 3, 1
branes which are the states of type IIB as we have carried out an
odd number (nine) of T-duality transformations. Notice that this
follows from the T-duality symmetry for fundamental strings, not
from T-duality arguments applied to
the above brane gases which we actually want to show.
Our demonstration is
done by carrying out explicitly nine T-duality transformations on
a mass state in $\cB$, and showing that there is a corresponding
and equal mass state in $\cB^*$. In this sense the two  brane
gases are T-dual.

Suppose that the branes in $\cB$ are wrapped around some of the
cycles of a nine-torus with radii $(R_1, \cdots, R_9)$. Then, the
volume $Vol_p$ of a p-brane in Eq. (\ref{eq:Ep}) is simply the
product of the p circumferences, and the minimal masses $M_p =
E_p$ (in the string frame) are
\begin{eqnarray}
  &&  M_{0}=\frac{1}{g \al^{'1/2}} \label{eq:M0}\,,\\
  &&  M_{2}=(2\pi)^2  R_9 R_8 \tau_2 = \frac{R_9 R_8}{g \al^{'3/2}}\,,\label{eq:M2}\\
  &&  M_{4}=(2\pi)^4  R_9 R_8 R_7 R_6 \tau_4 = \frac{R_9 R_8 R_7 R_6}{g \al^{'5/2}}\,,\\
  &&  M_{6}=(2\pi)^6  R_9 \cdots R_4 \tau_6 = \frac{R_9 \cdots R_4}{g \al^{'7/2}}\,,\\
  &&  M_{8}=(2\pi)^8  R_9 \cdots R_2 \tau_8 = \frac{R_9 \cdots R_2}{g
            \al^{'9/2}}\label{eq:M8} \, ,
\end{eqnarray}
(see also \cite{Polchinski:1998rr}) where in the second step we
have used expression (\ref{eq:tension}) for the tension of a
p-brane. For notational convenience we have fixed a
par\-ti\-cu\-lar winding configuration. The argument is
generalized for arbitrary winding configurations and winding
numbers at the end of this section. If, as we have assumed, $R_n
> \al^{' 1/2}$, then the heaviest object in the theory is the
8-brane.

The DBI action (\ref{eq:dbiaction}) is invariant under T-duality.
Hence, all formulae derived from it (energy, mass and pressure)
are valid in both the original brane gas $\cB$ and in the dual
brane gas $\cB^*$. Thus, the mass spectrum of the $\cB^*$ brane
gas is
\begin{eqnarray}
  &&  M^*_{9}=(2\pi)^9  R'_9 \cdots R'_1 \tau^*_9 = \frac{R'_9 \cdots R'_1}{g^* \al^{'10/2}}\label{eq:M9}\,, \\
  &&  M^*_{7}=(2\pi)^7  R'_7 \cdots R'_1 \tau^*_7 = \frac{R'_7 \cdots R'_1}{g^* \al^{'8/2}}\,,\\
  &&  M^*_{5}=(2\pi)^5  R'_5 \cdots R'_1 \tau^*_5 = \frac{R'_5 \cdots R'_1}{g^* \al^{'6/2}}\,,\\
  &&  M^*_{3}=(2\pi)^3  R'_3 R'_2 R'_1 \tau^*_3 = \frac{R'_3 R'_2 R'_1}{g^* \al^{'4/2}}\,,\\
  &&  M^*_{1}= 2\pi  R'_1 \tau^*_1 = \frac{R'_1}{g^* \al'}\,.
  \label{eq:M1}
\end{eqnarray}
Since $R_i^{'} < \al^{' 1/2}$, the heaviest brane of the dual gas
$\cB^*$ is now the 1-brane. The coupling constant in $\cB^*$ is
given by
\begin{equation} \label{cctrans}
    g^*=\frac{\al^{'9/2}}{R_9 \cdots R_1}g\,.
\end{equation}
Note that if the radii $(R_1,\cdots,R_9)$ of the initial
nine-torus are bigger than the self-dual radius $\al^{'1/2}$, then
$g^*<g$, and thus the assumption of a small string coupling
constant is safe.

Given the two mass spectra, one can easily verify that each state
in the brane gas $\cB$ has a corresponding state with equal mass
in the dual brane gas $\cB^*$:
\begin{equation}
    M^*_{9-p} = M_p \, .
\end{equation}
This establishes explicitly that the T-duality of the string gas
used in \cite{Brandenberger:1989aj} extends to the brane gas
cosmology of \cite{Alexander:2000xv}.

As an explicit example, consider a 2-brane wrapped around the 8
and 9 directions. Its mass is (\ref{eq:M2})
\begin{equation}
M_{2} = \frac{R_8 R_9}{g \al^{'3/2}} \, .
\end{equation}
If we replace the string coupling constant $g$ by the dual string
coupling constant $g^*$ via (\ref{cctrans}), and the radii $R_8$
and $R_9$ by the dual radii $R_8^{'}$ and $R_9^{'}$ via
(\ref{eq:tdualitydef}), one obtains
\begin{equation}
M_{2} = \frac{R_1^{'} \cdots R_7^{'}}{g^* \al^{' 8/2}} = M_{7}^*
\, .
\end{equation}

In the above example, we have specified a particular winding
configuration for simplicity, but clearly the argument holds as
well in the general case, where a p-brane wraps around some
directions $n_1 \cdots n_p$:
\begin{equation}
    M_{p}=(2\pi)^p R_{n_1} \cdots R_{n_p} \tau_p = \frac{R_{n_1}
    \cdots R_{n_p}}{g \al^{'(p+1)/2}} \, .
\end{equation}
Via the same steps as in the above example, it follows that
\begin{eqnarray}
    M^*_{9-p} &=& (2\pi)^{9-p} R'_{m_1} \cdots R'_{m_{9-p}} \tau^*_{9-p}
    = \frac{R'_{m_1} \cdots R'_{m_{9-p}}}{g^* \al^{'(10-p)/2}} \nonumber \\
              &=& M_p \, ,
\end{eqnarray}
where $\{m_1, \cdots, m_{9-p}\} \neq \{n_1, \cdots, n_{p}\}$.


\subsection{Multiple windings}

Consider now a p-brane with multiple windings
$\om=(\om_1,\cdots,\om_p,0,\cdots,0)$. Its mass is
\begin{equation}
    M_p(\om) = \om_1 \cdots \om_p M_p\,.
\end{equation}
In the $\cB^*$ brane gas this corresponds to a number $\om_1
\cdots \om_p$ of (9-p)-branes each with winding
$\om^*=(0,\cdots,0,1,\cdots,1)$ and mass $M^*_{9-p}$. Since
$M^*_{9-p}=M_p$, the total mass of this `multi-brane' state is
equal to the mass of the original brane, namely
\begin{equation}
     (\om_1 \cdots \om_p) M^*_{9-p}=M_p(\om) \,,
\end{equation}
which establishes the correspondence of $\cB$ and $\cB^*$ in the
case of multiple windings.

One should also add fundamental strings to the brane gas $\cB$.
Since their mass squared is
\begin{equation}
    M^2=\left(\frac{n_n}{R_n}\right)^2+\left(\frac{\om_n
    R_n}{\al'}\right)^2\,,
\end{equation}
is its clear that every fundamental string state in $\cB$ has a
corresponding state in $\cB^*$ when $n_n \leftrightarrow \om_n$.


\section{Cosmological Implications and Discussion}
\label{sec:cosmology}

We have demonstrated explicitly how T-duality acts on a brane gas
in a toroidal cosmological background, and have in particular
shown that the mass spectrum of the theory is invariant under
T-duality. Thus, the arguments of \cite{Brandenberger:1989aj}
which led to the conclusion that cosmological singularities can be
avoided in string cosmology extend to brane gas cosmology.

Whereas T-duality does not change the nature of fundamental
strings, but simply interchanges winding and momentum numbers, it
changes the nature of branes: after T-dualizing in all $d$ spatial
dimensions, a $p$-brane becomes a $(d-p)$-brane which, however,
was shown to have the same mass as the `original' brane.

In \cite{Alexander:2000xv} it was shown that the dynamical
decompactification mechanism proposed in
\cite{Brandenberger:1989aj} remains valid if, in addition to
fundamental strings, the degrees of freedom of type IIA
superstring theory are enclosed. We briefly comment on the
decompactification mechanism in the presence of a type IIB brane
gas on a nine-torus. As before, we assume a hot, dense initial
state where, in particular, the brane winding modes are excited
and in thermal equilibrium with the other degrees of freedom. All
directions of the torus are roughly of string scale size, $l_s$,
and start to expand isotropically. For the 9-,7, and 5-brane
winding modes there is no dimensional obstruction to continuously
meet and to remain in equilibrium, thereby transferring their
energy to less costly momentum or oscillatory modes: these degrees
of freedom do not constrain the number of expanding dimensions.
However, the 3-branes allow only seven dimensions to grow further,
and out of these, three dimensions can become large  when
1-brane and string winding modes have disappeard.
As far as the
`intercommutation' and equilibration process is concerned, the
1-branes and the strings play the same role, but since the winding
modes of the former are heavier ($\frac{R}{g\al'} \gg
\frac{R}{\al'}$ at weak coupling), they disappear earlier.

We have focused our attention on how T-duality acts on brane
winding modes. However, since in a hot and dense initial state we
expect all degrees of freedom of a brane to be excited, we should
also include transverse fluctuations (oscillatory modes) in our
considerations concerning T-duality. To our knowledge, the
quantization of such modes is, however, not yet understood, and we
leave this point for future studies.

Another interesting issue is to investigate how the present
picture of brane gas cosmology gets modified when gauge fields on
the branes are included. These correspond to $U(N)$ Chan-Paton
factors at the open string ends. In this case, a T-duality in a
transverse direction yields a number $N$ of parallel (p-1)-branes
at different positions \cite{Polchinski:1998rq}.



\section*{Acknowledgments}

We would like to thank S. Alexander, L. Alvarez-Gaume, J.
 Fernando-Barbon, S. Foffa, S. Lelli, J. Mourad, R. Myers, Y. Oz, F. Quevedo,
A. Rissone, M. Rozali and M. Vasquez-Mozo  for useful discussions. R.B.
wishes to thank the CERN Theory Division and the Institut
d'Astrophysique de Paris for their hospitality and support during
the time the work on this project was done.  He also acknowledges
partial support from the US Department of Energy under Contract
DE-FG02-91ER40688, TASK A.


\section*{Appendix}

In this appendix we would like to give some further arguments for why we
believe brane gas cosmology to be non-singular. First, let us recall
the results of \cite{Brandenberger:1989aj}. In the case of a string
gas, where the strings are freely propagating on a 9-torus, it was
shown that the 
cosmological evolution is free of singularities. The background
space-time has to be compact, otherwise the thermodynamical
description of strings is not sound, in particular the specific heat
becomes ne\-ga\-tive at large energies. An important assumption in the
derivation was that the evolution of the universe is
adiabatic, i.e. the entropy of the string gas is constant. Making use
of this assumption,
one can find the temperature as a function of the scale factor,
$T(R)$, without referring to the dynamics of gravity or Einstein's
equations. Using a microcanonical approach, it was shown that there
exists a maximum temperature, called Hagedorn temperature
$T_H$, and hence there is no temperature singularity in string gas
cosmology. The curve $T(R)$ is invariant under a T-duality
transformation, $T(R)=T \left(\alpha'/R \right)$. Another crucial
assumption is that the string coupling constant, $g$, is small enough
such that the thermodynamical computations for free strings are
applicable, and that the back-reaction of the string condensate
on the background geometry can be neglected. In lack of knowledge
about brane thermodynamics we simply postulate that the statements
above extend to brane gases. 

We conclude by making a comment on our work in the light of the
well-known singularity theorems in General Relativity. These theorems
make assumptions about the geometry of space-time such as 
$R_{\mu\nu}\xi^{\mu}\xi^{\nu} \geq 0$ for all timelike vectors
$\xi^{\mu}$ (for a textbook treatment see e.g. \cite{Wald:1984rg}). 
By Einstein's equations this is equivalent to the strong
energy condition for matter. However, we do not trust
Einstein's equations in the very early universe as they receive
corrections which are higher order in $\alpha'$ as well as $g$, and
also they lack invariance under T-duality transformations $R\rightarrow
1/R$. Therefore we cannot invoke the energy momentum tensor given in Eqs. 
(\ref{eq:finalrho}),(\ref{eq:0n}), (\ref{eq:trace}) to decide whether
the universe described by our model is geodesically complete or not.


\end{document}